\documentclass[]{aa}  

\usepackage{diagbox}
\usepackage{graphicx}
\usepackage{natbib}
\usepackage{hyperref}
\usepackage{txfonts}

\DeclareRobustCommand{\ION}[2]{%
\relax\ifmmode
\ifx\testbx\f@series
{\mathbf{#1\,\mathsc{#2}}}\else
{\mathrm{#1\,\mathsc{#2}}}\fi
\else\textup{#1\,{\mdseries\textsc{#2}}}%
\fi}

\newcommand{\nii}{[\ION{N}{ii}]}

\newcommand{\oiii}{[\ION{O}{iii}]}

\newcommand{\Ha}{$\rm{H}\alpha$}
\newcommand{\Hb}{$\rm{H}\beta$}

\newcommand{\kms}{km\,s$^{-1}$}

\newcommand{\edr}{S\'anchez et al. (submitted)}

\def\pyp{\texttt{pyPipe3D}\xspace}

\begin{document}

   \title{The WHaD diagram: Classifying the ionizing source with one single emission line}
   

   \author{S.~F.~S\'anchez\inst{\ref{unam}}
        \and A.~Z.~Lugo-Aranda\inst{\ref{unam}}
        \and J. S\'anchez Almeida\inst{\ref{iac},\ref{ll}}
        \and J.~K.~Barrera-Ballesteros\inst{\ref{unam}}
        \and O.~Gonzalez-Martín\inst{{\ref{irya}}}
        \and S.~Salim\inst{\ref{indi}}
        \and C.~J.~Agostino\inst{\ref{indi}}
 }

 \institute{Instituto de Astronom\'ia, Universidad Nacional Auton\'oma de M\'exico, A.P. 70-264, 04510, M\'exico, CDMX \label{unam}
 \and Instituto de Astrof\'\i sica de Canarias, La Laguna, Tenerife, E-38200, Spain \label{iac}
 \and Instituto de Radioastronom\'\i a and Astrof\'\i sica (IRyA-UNAM), 3-72 (Xangari), 8701, Morelia, Mexico \label{irya}
 \and Departamento de Astrof\'\i sica, Universidad de La Laguna, Spain \label{ll}
 \and Department of Astronomy, Indiana University, Bloomington, IN 47405, USA \label{indi}
}


   \date{Received ---, 2023; accepted ---, 2023}


  \abstract
  {The usual approach to classify the ionizing source using optical spectroscopy is based on the use of diagnostic diagrams that compares the relative strength of pairs of collisitional metallic lines (e.g., \oiii\ and \nii) with respect to recombination hydrogen lines (e.g., \Hb\ and \Ha). Despite of being accepted as the standard procedure, it present known problems, including confusion regimes and/or limitations related to the required signal-to-noise of the involved emission lines. These problems affect not only our intrinsic understanding of inter-stellar medium and its poroperties, but also fundamental galaxy properties, such as the star-formation rate and the oxygen abundance, and key questions just as the fraction of active galactic nuclei, among several others}
   {We attempt to minimize the problems introduced by the use of these diagrams, in particular their implementation when the available information is limited, either because not all lines are available or the do not have the required signal-to-noise.}
   {We explore the existing alternatives in the literature to minimize the confusion among different ionizing sources and proposed a new simple diagram that uses the equivalent width and the velocity dispersion from one single emission line, H$\alpha$, to classify the ionizing sources.}
   {We use aperture limited and spatial resolved spectroscopic data { in the nearby Universe (z$\sim$0.01) } to demonstrate that the new diagram, that we called WHaD, segregates the different ionizing sources in a more efficient way that previously adopted procedures. A new set of regions are defined in this diagram to select betweeen different ionizing sources.}
   {{ The new proposed diagram is well placed to determine the ionizing source when only H$\alpha$ is available, or when the signal-to-noise of the emission lines are too low to obtain reliable fluxes for the weakest emission lines involved in the classical diagnostic diagrams (e.g., H$\beta$).}}

   \maketitle
%
\section{Introduction}
\label{sec:intro}

Understanding of which sources ionize the interstellar medium (ISM) in
galaxies is of a paramount importance for the exploration not only of
the properties of the ISM itself, but also for a correct evaluation of
fundamental parameters, such as the star-formation rate or the metal
abundance, that trace the current stage and past evolution of these
objects \citep[e.g., ][for recent reviews on the
topic]{kewley19,sanchez21}. Classically the dominant ionizing source,
both galaxy wide or in a region within a galaxy, has been determined
using the so-called diagnostic diagrams when using optical
spectroscopy. These diagrams compare the distributions of different
ionizing sources for a set of pairs of line ratios between
collisionally excited emission lines (e.g., \oiii 5007 and \nii 6584)
and the nearest (in wavelength) hydrogen recombination line (e.g. \Hb
and \Ha) \citep[e.g.][]{baldwin81,osterbrock89,veil01}.

The most frequenly used of those diagrams, usually known as the BPT
diagram \citep[BPT][see, Fig. \ref{fig:diag}, left panel]{baldwin81}, compares the
distribution of the logarithm of the \oiii/\Hb\ line ratio (O3) as a
function of the logarithm of the \nii\/\Ha\ one (N2).  In this diagram
the ionization due to young-massive OB stars, those born in recent
star-formation episodes, follows a well defined arc-shape, with
metal-rich ionizing stars at the bottom-right end of that arc and
metal-poor ones at the top-left extreme. This arc is the result of the
limited range of shapes of ionizing spectra for this kind of sources
(being just a certain stellar type modulated by its metal content). On
the contrary, ionizing sources with harder UV-spetra produce different
line ratios, that could be in principal above the loci traced by
OB-stars \citep[][]{osterbrock89}. Based on this physical principle,
different demarcation lines have been proposed to classify the
ionizing source. Among them, the most frequently used ones are those
proposed by \citet{kauff03} and \citet{kewley01} to distiguish between
ionization related to recent star-formation (SF) processes and those
related to active galactic nuclei (AGN). Despite the fact that they
have a very different origin and they are based on different
assumptions \citep[read Sec.4 in ][ for a detailed discussion on the
topic]{sanchez21}, they are frequently used in combination to define
three regimes in the O3-N2 plane, associated with three ionizing
sources: (i) SF, below the \citet{kauff03} demarcation line, (ii)
AGNs, above the \citet{kewley01} one, and (iii) mixed/intermediate,
for values within both lines\footnote{The list of references using
  this scheme is too large to cite them here}.

This approach is often far too simplistic and strongly biased by the
overwhelming number of explorations using single aperture
spectroscopic surveys \citep[e.g., SDSS or GAMA][]{york2000,gamma}. First,
the so-called intermediate regime could be populated by ionization
purely related to a recent SF event, in particular in the presence of
super-nova remnants \citep[e.g.][]{cid21}, or by ionization due to
low-metallicity AGNs. This is indeed confirmed when exploring the
location of X-ray selected AGNs in this diagram \citep[X-AGNs,
e.g.,][]{agos19,nata23,agos23} Thus, the concept of intermediate/mixed
region is often misleading, as least in general, despite of the fact
that this regime could indeed be populated by the mix of SF an harder
ionization sources \citep{davies16,lacerda18}.

Finally, it is well known that other ionizing sources can mimic the
line ratios usually associated to just AGNs when using this
classification scheme. For instance, shocks presents a wide variety of
line ratios depending on the properties of the ionized gas, the
strength of the magnetic field and the velocity of the
shock. High-velocity ones, those associated with galaxy-scale outflows
\citep[e.g.][]{carlos20}, have line ratios similar to those observed
in strong AGNs \citep[e.g.][]{veil01}. On the contrary, low-velocity
ones, observed for instance in certain elliptical galaxies, present
much lower line ratios \citep[e.g.][]{dopita96}.  Another key ionizing
source in this kind of galaxies, and in non-SF retired galaxies in
general \citep[RG,][]{sta08}, or retired regions within galaxies
\citep[e.g.][]{singh13,belfiore17a,lacerda18}, is the one produced by
Hot evolved and post-AGB stars \citep{binette94}.  The ionization by this stars produces
weaker emission lines than those due to SF or AGNs (and shocks in many
cases). However their line ratios cover a wide range of values, being
as low as the values usually associated with SF or as high as the
those classically assigned to AGNs \citep{ARAA}. We will refer to this
ionization as RG along this article .

The distribution in the O3-N2 BPT diagram of both shocks and post-AGB
ionization is very similar \citep[e.g.][]{sanchez21}. They cover the
regime from the bottom-end of the SF distribution expanding towards
larger values of both line ratios until reaching the same regime
covered strong AGNs. They can also cover the same area known as
low-ionization nuclear emission region \citep[LINER][]{heckman87},
that nowadays is not consider to be produced by low-luminosity AGNs
in all the cases. Therefore, the line ratios adopted by the traditional
BPT are not sufficient to separate between certain ionizing
sources. To circumvent this problem, there have been proposed
different strategies. For instance, \citet{cid-fernandes10} introduced
the WHaN diagram, using the equivalent-width of H$\alpha$,
EW(H$\alpha$) (or WH$\alpha$), a tracer of the relative intensity of
the emission lines with respect to the underlying continuum, and the
N2 ratio, as a tracer of the hardness of the ionization. In this
diagram SF and AGNs are consider to have large values of
EW(H$\alpha$), with the former having lower values of N2 that the
later. Regions ionized by hot evolved stars, present low
values of EW(H$\alpha$) ($<$3\AA), irrespectively of the N2
value. Based on a similar reasoning, \citet{sanchez14} proposed to
combine the O3-N2 BPT diagram with the EW(H$\alpha$) to clean the SF
regime from the contamination by low-N2 RG sources. Later
\citet{sanchez18} generalized this idea, introducing the
BPT+WHa diagram, in which the ionization is classified using both the
BPT diagram and the EW(H$\alpha$) for all the considered ionizing
sources.

The use of EW(H$\alpha$) helps to distiguish between RG
and both SF and AGNs where they overlap in the BPT diagram.
However, it does not completely solve the problem of identifying the
dominant ionizing source. For instance, in the case of shocks, both
the location in the BPT diagram and the value of the EW(H$\alpha$)
could be simular to that observed for AGNs (and SF in some cases).  A
possible solution is to explore the shape of the ionized structure,
what it is only possible when spatial resolved spectroscopic data is
available \citep[e.g.][]{jarvis90,carlos17,carlos19}. An alternative
strategy is to introduce the asymmetries in the emission lines and the
velocity dispersion as an additional parameter to
distiguish between different ionizing sources, or a combination of
those kinematics parameters with the BPT, the location within the
galaxy and/or the galactocentric distance
\citep{dagos19,carlos20,john23}.

The methods including velocity information are indeed physical 
motivated: (1) SF happens in disks, and it requires low velocity
dispersions to be triggered; (2) RG is associated with hot-evolves
stars, that in general are found in bulges or thick disks, i.e., they
are associated with regions of larger velocity dispersions than disks;
(3) AGN ionization happens in the so called narrow-line regions
\footnote{for type-II AGNs, as type-I are the more easy ones to be
  identified by their distinguish feature of having broad Balmer
  lines}, that, despite of its name, presents much larger velocity
dispersions than galaxy disks; and finally (4) shock ionization, by its
very nature is associated with high-velocity dispersion clouds, due to
the result of the compression and non uniform propagation of the
galactic winds (shocks also produce multi-component and asymmetric
emission lines). A recent exploration by \citet{law21} demonstrates
that SF ionization is statisitically associated with low velocity
dispersion emission lines ($\sigma_{\rm vel}\sim$25\kms), while harder
ionizations are associated wtih a wide range of velocity dispersions,
larger than $>$50-60 \kms in general. Indeed, the average distribution
of the ionized gas velocity dispersions across the BPT diagram follows
a similar, but oposite, pattern of that described by the EW(H$\alpha$)
\citep[e.g.][]{lacerda18,sanchez20,sanchez22}.


Often the methods described above are insufficient. Determining the
nature of the ionization requires the use of additional information,
such as the shape of the ionized structures, the fraction of young
stellar populations, the location with in the galaxy and the
galactocentric
distance\citep[e.g.][]{dagos19,carlos20,espi20,sanchez21,john22}.  The
more additional parameters are used the more complicated is to observe
all of them simultaneously with the required signal-to-noise to
provide reliable classifications.  Diagnostic diagrams based
on line ratios require the simultaneous determination of the flux
of four emission lines, not all of them of the same
relative intensity (and  signal-to-noise). For instance, in the case of
the O3-N2 diagram it is often the case that H$\beta$ is the
limiting parameter. Ionizing sources that generate intrinsically weak
emission lines, such as post-AGB/hot evolved stars, produce H$\alpha$
fluxes barely detected with the current spectroscopic galaxy surveys
in many cases. In those cases H$\beta$, $\sim$3 times weaker, remains
frequently undetected. A similar situation may happens in the presence
of strong dust attenuation, like in the case of Ultra-Luminous
Infrared galaxies, despite of the intrinsic luminosity of the ionized
gas emission.  As expected the situation becomes more
complicated when more parameters are introduced. So, the seek for a
more accurate classification is often hampered by the precision at
which the involved parameters could be estimated.

One additional problem of most classical diagnostic diagrams like the
BPT is the use of lines appearing in several different spectral
regions. Due to technical limitations or the design of the
spectrograph, the lines cannot be observed simultaneously in a single
setup \citep[e.g., MEGARA][]{megara} or even with a single
spectrograph \citep[MUSE][and the need for BlueMUSE]{muse}.
These problems disappear if there were diagnostics based
on a single line.


Motivated by the limitations of the currently adopted procedures to
classify the ionizing source, and based on the most recent
explorations that consider both the EW(H$\alpha$) and
$\sigma_{\rm vel}$ as key parameters in this regards, we present here
a single emission line diagnostic diagram that uses only those two
parameters (the WHaD hereafter). We demonstrate that this diagram
determines the nature of the ionizing source as well as the combination of
the BPT diagram with the EW(H$\alpha$), with the advantage of using
just one single line and two parameters. We illustrate its
use by comparing the spatial resolved classification of the ionizing
source provided by the WHaD diagram with that provided by the BPT+WHa
for a few archetypal galaxies, providing access to a similar comparison
for a large galaxy sample.

The distribution of this article is as follows: the different datasets
employed in this study are presented in Sec. \ref{sec:data}, with both
the analysis and results included in Sec. \ref{sec:ana}. Finally, the
main conclusions are summarized in Sec. \ref{sec:con}.

\section{Data}
\label{sec:data}

We make use of the dataproducts provided by the \pyp pipeline
\citep{pypipe3d} for the integral field spectroscopy (IFS) data
provided by the eCALIFA (extended Calar Alto Legacy Integral Field
Area) survey \citep[][, Sanchez et al., submitted]{califa}, and the
final distribution of the MaNGA surveys \citep[Mapping Nearby Galaxies
at APO][]{manga}. Both surveys have been extensively described in the
literature, in particular in the articles describing their most recent
and final data releases \citep{sanchez16,DR17} and \edr, with a
comparison between them included in a recent review on the topic of
IFS galaxy surveys \citep{ARAA}.

To avoid repetitions we will not include a throughfull description of
those surveys in here. For the purposes of the current exploration, we
should highlight that: (i) both surveys observe a representative
sample of galaxies in the nearby universe ($z_{\rm CALIFA}\sim 0.015$,
$z_{\rm MaNGA}\sim0.045$), coverig a wide range of galaxy types and
stellar masses, being complete above
M$_\star$$\sim$10$^{9}$M$_\odot$, once applied the proper volume
correction; (ii) the sample of galaxies is
$\sim$10 times larger in the case of MaNGA
($\sim$9000 objects with good quality data) with respect to the sample
covered by eCALIFA (895 objects), but the field-of-view of the IFS
data (74$\arcsec$ vs.
19-32$\arcsec$), the spatial coverage (2.5 Re vs. 1.5-2.5 Re),
sampling and physical resolution ($\sim$0.3 kpc vs
$\sim$2 kpc) is better/larger in the case of eCALIFA (in average);
(iii) the spectral range (resolution) of the MaNGA data is
$\sim$2 times larger (better) than that of eCALIFA,
R$\sim$2000 vs
$\sim$850 respectively. Nevertheless, in both cases is good enough for
the purposes of this study; and finally (iv) both surveys cover the
emission lines of interest for the current exploration (\Hb, \oiii,
\Ha\ and \nii), with enough signal-to-noise to measure the flux
intensities, velocity dispersion and equivalenth widths for the vast
majority of the observed objects.

\pyp is an automatic pipeline that performs a spectral decomposition
of the stellar continuum and the ionized gas emission lines based on
the stellar synthesis tecnique, assuming a particular template of
single stellar populations \citep{pipe3d_ii,pypipe3d}. This pipeline
has been extensively used to study IFS data from
different datasets, including CALIFA
\citep[e.g.][]{mariana16,espi20}, MaNGA
\citep[e.g.][]{ibarra16,jkbb18,sanchez18b,bluck19,laura19}, SAMI
\citep[e.g.][]{sanchez19}, and AMUSING++ 
\citep{laura18,carlos20}. In addition, it has been tested using
mock datasets based on hydrodynamical simulations \citep{guidi18,ibarra19,sarm23}.
To avoid unnecessary repetition we do not include a description of the
\pyp\ either.

For the purpose, of the current exploration it is just relevant to
know that for each analyzed datacube \pyp provides the spatial
distribution of a set of physical and observational parameters derived
for both the stellar population and the ionized gas. In this study we
make use of the spatial distributions of (i) the flux of the emission
lines involved in the O3-N2 BPT diagram, (ii) the H$\alpha$ velocity
dispersion (corrected by instrumental resolution), and (iv) the
EW(H$\alpha$), all of them recovered based on the non-parametric
moment analysis peformed by this pipeline on the pure gas datacubes
(i.e., the datacube once subtracted the best model spectra for the
stellar component). All these dataproducts are publicly available
distributed by \edr\ and \citet{sanchez22}, as part of the
corresponding eCALIFA and MaNGA data releases. It was also distributed
the average value of each of these parameters at both the effective
radius (Re) and in the central region of each galaxy (1.5$\arcsec$
diameter and 2.5$\arcsec$ diameter for eCALIFA and MaNGA,
respectively).


In addition, we make use of the results by \citet{nata23}, which have
recently explored the X-ray properties of a sub-sample of the CALIFA
galaxies, recovering $\sim$40 bona-fide AGNs. Finally, we use the O3
and N2 values of the sample of X-ray selected AGNs presented in
\citet{agos19} for comparison purposes.

\section{Analysis and Results}
\label{sec:ana}

   \begin{figure*}
   \centering
   \minipage{0.99\textwidth}
   \includegraphics[width=6cm,trim={20 5 20 0},clip]{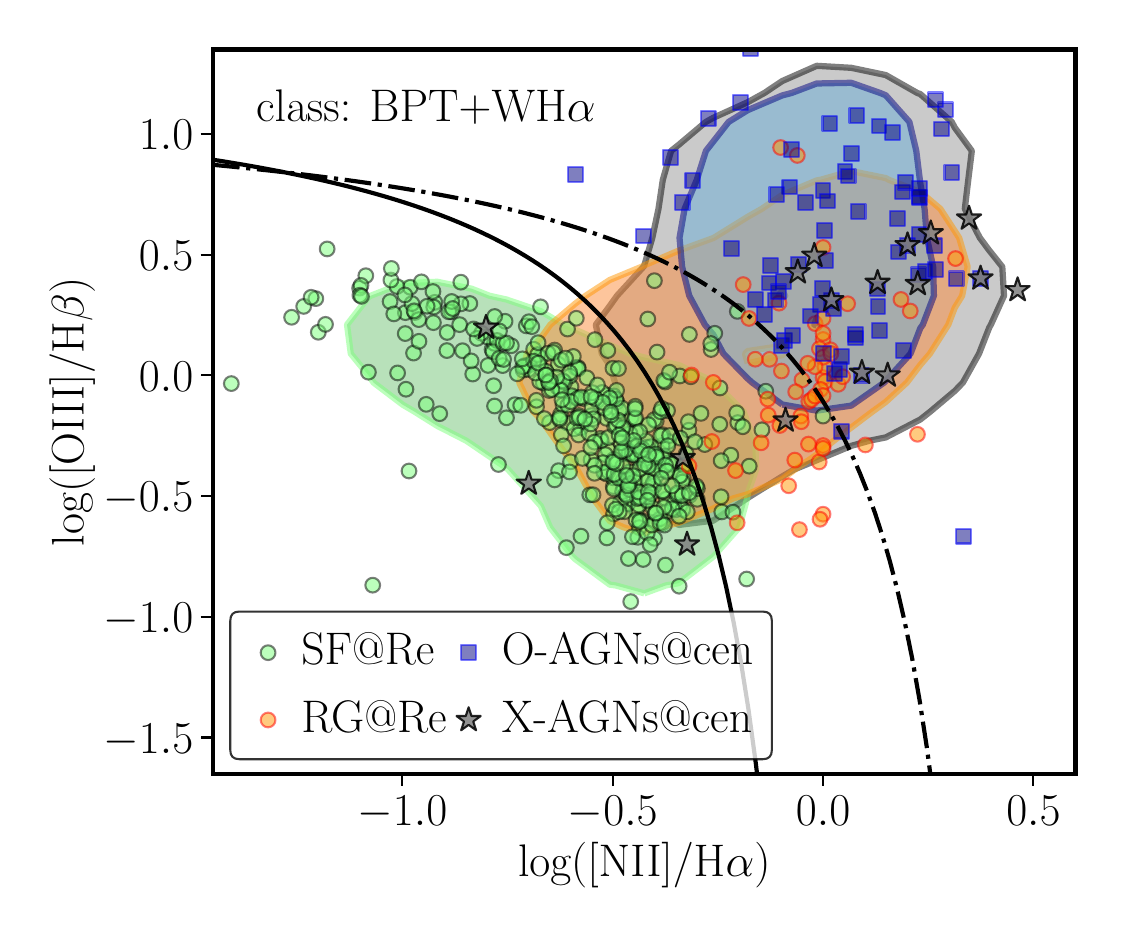}
   \includegraphics[width=6cm,trim={20 5 20 0},clip]{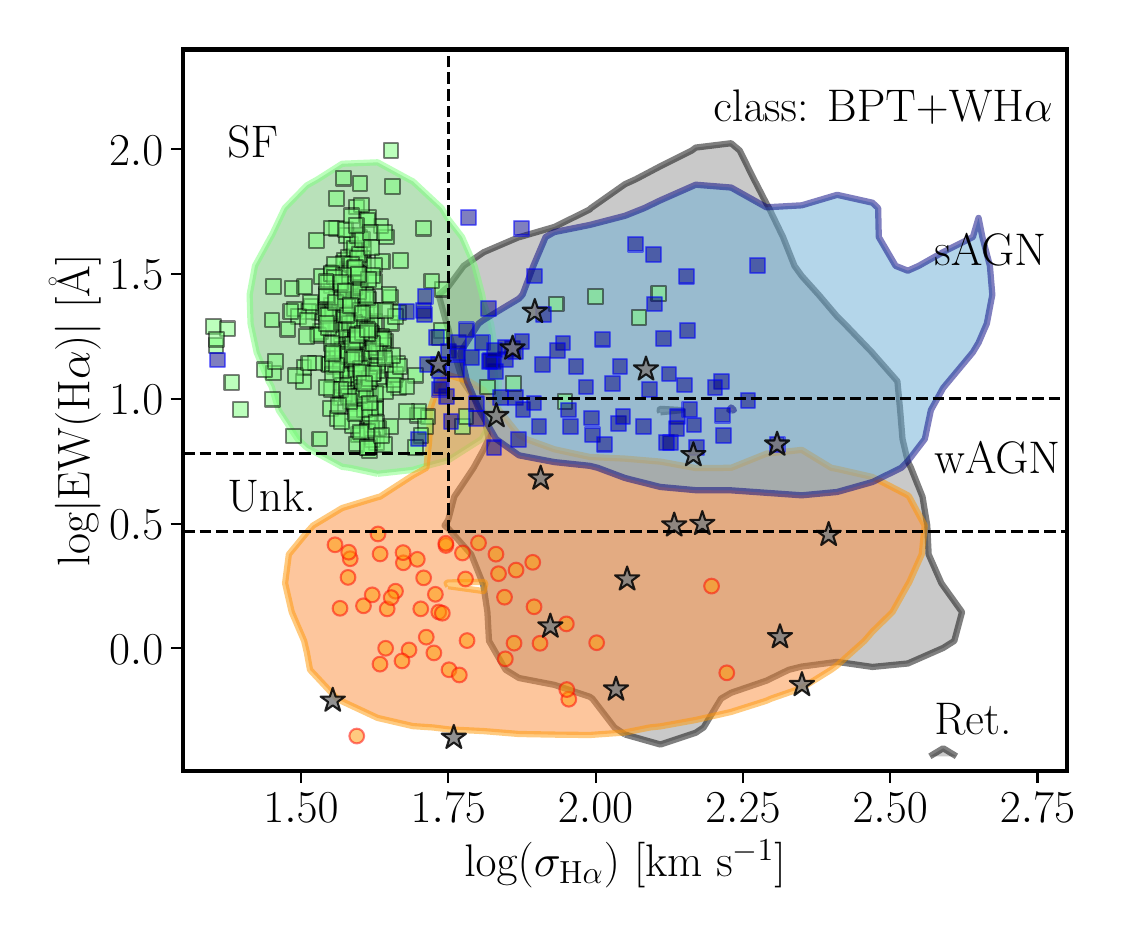}
 \includegraphics[width=6cm,trim={20 5 20 0},clip]{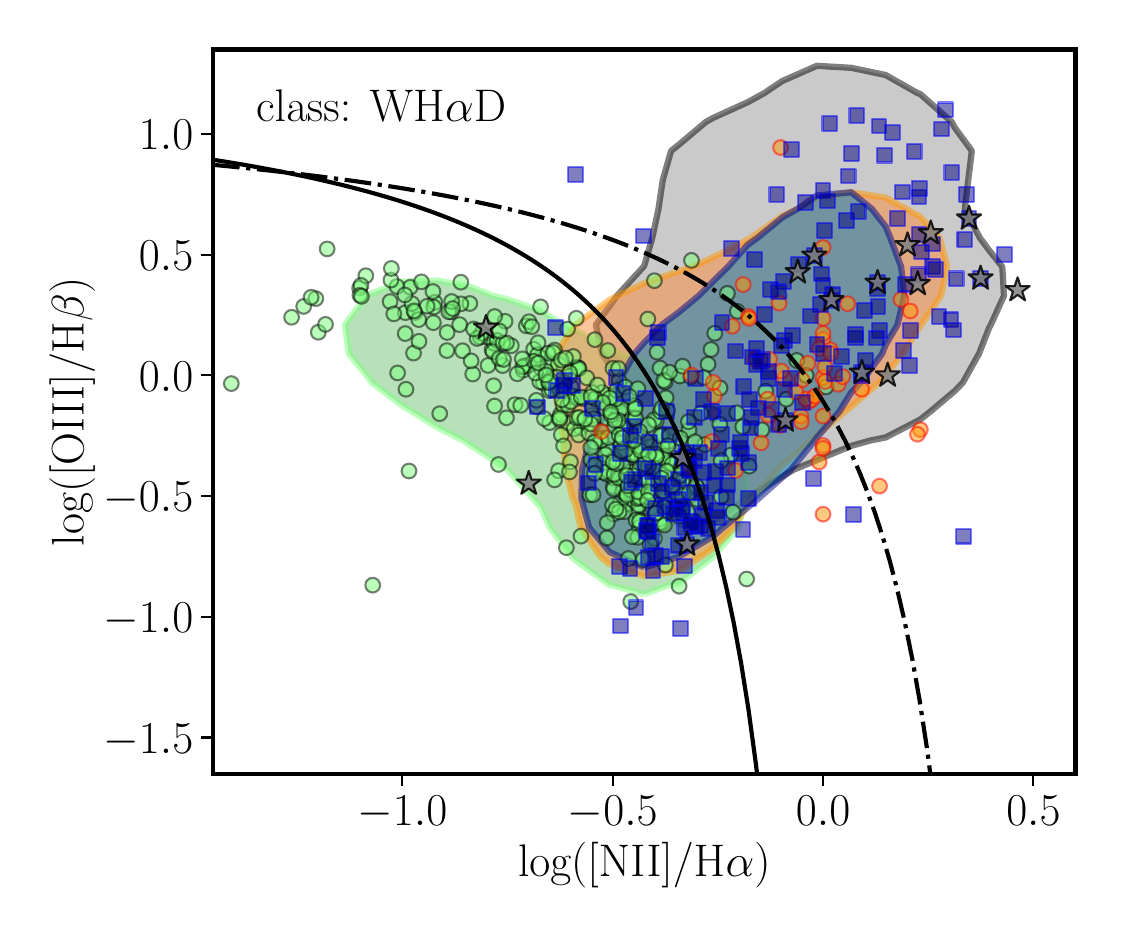}
 \endminipage
 \caption{{\it Left panel:} classical BPT diagram \citep{baldwin81}, showing the distribution of \oiii/H$\beta$ line ratio as a function of  \nii/H$\alpha$ ratio. It includes four sub-samples of objects whose ionization has been classified according to this diagram together togheter with  EW(H$\alpha$), following \citet{sanchez21}: (i) star-forming galaxies (SF, green), (ii) retired galaxies (RG, orange), (iii) optically selected AGNs (O-AGNs, blue) and (iv) X-ray selected AGNs (X-AGNs, black stars). Solid symbols correspond to data extracted from the eCALIFA sample, with the X-AGNs being extracted from \citet{nata23}. The shadded regions correspond to data extracted from the MaNGA sample (SF, RG and O-AGNs) and from \citet{agos19} (X-AGNs), showing the area encircling 90\%\ of the objects in each case. The region at which the ionization was measured is indicated in the legend: central aperture (cen) or at the effetive radius (Re). Solid and dot-dashed lines correspond to the classical demarcations lines proposed by \citet{kauff03} and \citet{kewley01} to distinguish between the different ionizing sources. {\it Middle panel:} new proposed diagnostic diagram (WHaD) showing the distribution of the EW(H$\alpha$) (or WH$\alpha$) as a function of the H$\alpha$ velocity dispersion for the same sub-samples of objects included in the previous panel, using the same nomenclature. It is clearly appreciated that the different ionizing sources are well separated in this new diagram. Dashed lines correspond to the proposed demarcation lines to separate between different dominant ionizing sources, with the corresponding sources indicated. {\it Right-panel:} distribution in the BPT diagram shown in the first panel of the eCALIFA (solid symbols) and MaNGA (shadded regions) galaxies classified according to the location of their ionization in the WHaD diagram, measured at the same two locations indicated adopted for the values shown in the first panel. The location of the X-ray selected AGNs has been included for completness.}\label{fig:diag}
    \end{figure*}

\subsection{Classifying the ionization using the classical BPT diagram}
\label{sec:old_BPT}

As indicated in the introduction the procedure most frequently used to
classify the ionization is the location of the O3 and N2 line ratios
in the BPT diagram, combined, in some cases, with an addtional
parameter, for instance the EW(H$\alpha$). Figure \ref{fig:diag},
left-panel, shows the distribution in this diagram of the different
datasets described in the previous section, classified following the
prescriptions outlined in \citet{sanchez21}, hereafter the
BPT+WH$\alpha$ scheme: (1) RG, galaxies with EW(H$\alpha$) at Re 
below 3$\AA$, irrespectively of the location of the line ratios in the
BPT diagram, not belonging to any of the other groups; (2) SF,
galaxies with EW(H$\alpha$)$>$6$\AA$ and line ratios below the
\citet{kewley01} demarcation line at Re, and not belonging to any of
the other groups; (3) O-AGNs, galaxies with a EW(H$\alpha$)$>$6$\AA$
and line ratios above the \citet{kewley01} in the central
aperture. Note that for O-AGNs it is frequently distinguished between
strong and weak AGNs adopting an EW(H$\alpha$)$\sim$10$\AA$ as the
boundary between both categories. For AGNs, we selected the central
aperture due to the dilution effect introduced by the ionized gas from
the host galaxy \citep[e.g.][]{sanchez18,nata23,alba23}. Finally, we
include the X-AGNs extracted from \cite{nata23} and
\citet{agos19,agos23}. Note that most objects are well classified
using this scheme, with just a few objects without a well defined dominant
ionizing source.


By construction the SF galaxies are located where it is expected. They
follow a clear arc-shaped distribution not only below the adopted
demarcation line \citep{kewley01}, but in most of the cases below the
more restrictive \citet{kauff03} one ($\sim$92\% of the objects). A
similar behavior is found for the O-AGNs, that by construction are
located above the \citet{kewley01} curve. Thus, SF can be
distinguished from O-AGNs using just the BPT diagram and the classical
demarcation lines. However, the situation is not that simple when RG
are included. It is clear that galaxies (and regions within them) with
low EW(H$\alpha$) are distributed in the right-branch of the BPT
diagram, covering a wide range of values. Some 17\% of them are below
the \citet{kauff03} curve, $\sim$54\% in the intermediate region
between that line and the \citet{kewley01} one, and $\sim$28\% above
that line. Thus, without introducing an additional parameter, in this
case the EW(H$\alpha$), it is not possible to distinguish between this
ionizing source and SF or  O-AGNs.

We should note that the current selection of O-AGNs does not guarantee
that all AGNs are selected. It is well known that, in many instances,
radio-galaxies do not present signatures of optical AGNs regarding
their line ratios and/or flux intensities \citep[e.g.,][ and
references therein]{comer20}.  Other {\it bona fide} AGNs, like those
detected in X-rays, present emission lines in the optical which ratios
are not above the demarcation lines adopted to select O-AGNs in many
cases.  We include in Fig. \ref{fig:diag}, left-panel, the
distribution of the 424 X-ray selected AGNs by \citet{agos19} and
\citet{agos23} (X-AGNs hereafter), with the emission lines obtained
from the SDSS spectroscopic data. It is clear that they are not
restricted to the same regime usually adopted to select O-AGNs,
covering a range of values more similar to the one observed for
RG. Indeed, in the case of this particular sample of X-AGNs, only 34\%
of them would be classifed as O-AGNs, 28\% as RGs and 20\% as SF
galaxies, using the BPT-WH$\alpha$ classification scheme
\citep{sanchez21}. This is usually interpreted as a dilution effect by
the contamination of other ionizing sources, such as circumnuclear or
host-galaxy star-formation, or obscuration of the optical emission.
However, a similar behavior is observed in the distribution of the
line ratios extracted from the central aperture ($<$1kpc) of the
X-AGNs selected from the CALIFA sample itself studied by
\citet{nata23}. In this case the fraction of objects classified as
O-AGNs is even lower (17\%), using the BPT-WH$\alpha$
scheme. Furthermore, \citet{agos21} demonstrated that once subtracted
the SF contribution to the integrated SDSS spectra of a sample of
Seyfert galaxies a large number of them are still found below the
\citet{kauff03} demarcation line. In summary, neither the BPT along
nor its use combined with the EW(H$\alpha$) guarantee a clean
separation of AGNs from other ionizing sources.


   \begin{figure*}
   \centering
   \minipage{0.99\textwidth}
   \includegraphics[width=\linewidth,trim={20 20 20 20},clip]{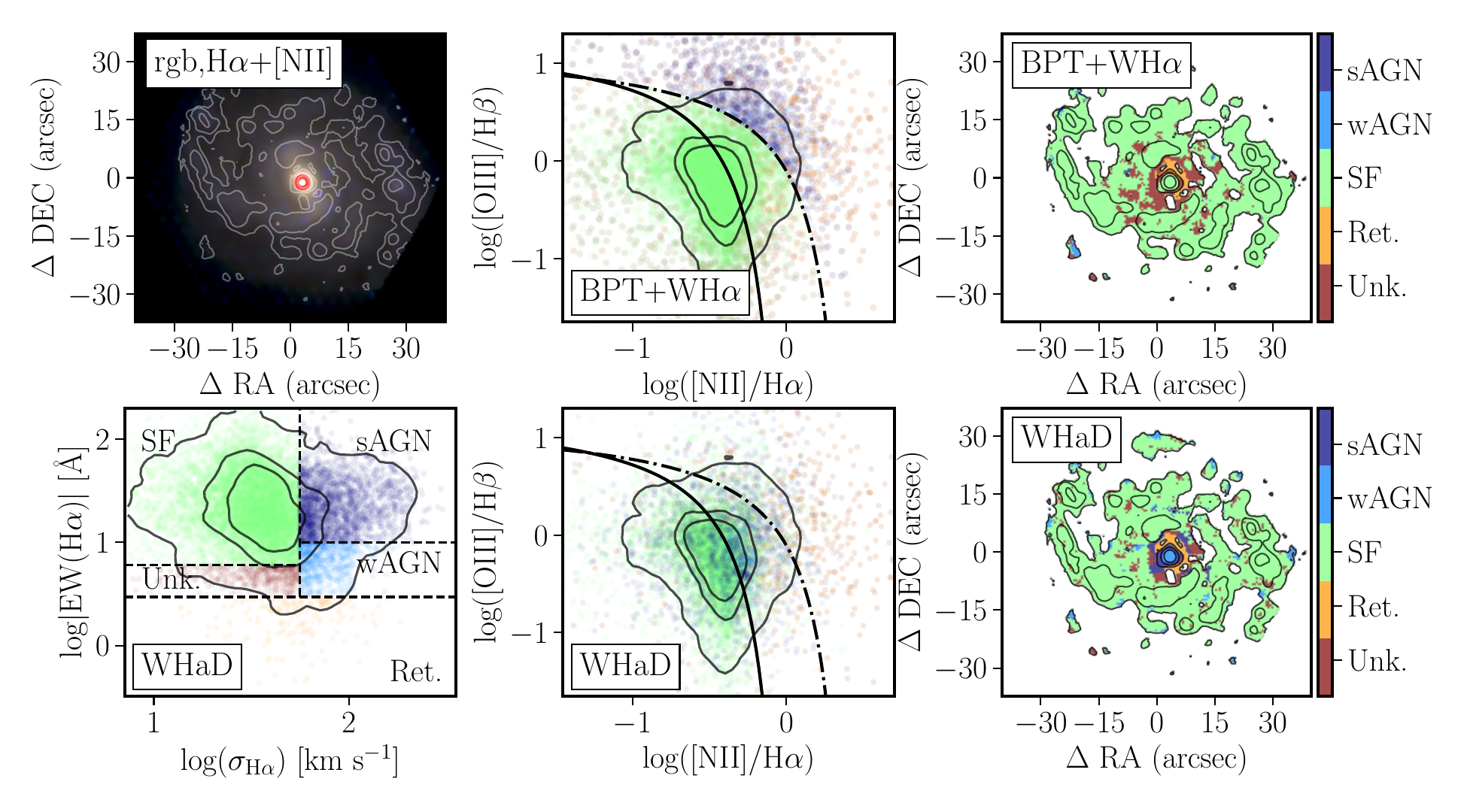}
 \endminipage
 \caption{{\it Top-left panel:} RGB-image created using the $r$-, $g$- and $r$-band images synthetized from the eCALIFA datacube of NGC\,5947, together with a set of contours showing the H$\alpha$+\nii\ flux intensities, with the 1st contour corresponding to the mean value minus one standard deviation, and each sucesive contour followig a multiplicative scale of five times that value. The red-circle indicates the central aperture adopted along this article (e.g., Fig. \ref{fig:diag}). {\it Top-middle panel:} classical diagnostic diagram showing the distribution of \oiii/H$\beta$ line ratio as a function of the \nii/H$\alpha$ ratio for each individual spaxel of the same cube shown in the top-left panel, color coded by nature of the ionization according to the classification based on locaton in this diagram together with the EW(H$\alpha$), thus, the same selection criteria adopted in top-left panel of Fig. \ref{fig:diag}: SF in green, RG in orange and AGN-like ionization in blue. {\it Top-right panel:} spatial distribution of the ionized gas for those spaxels with H$\alpha$ above the mean minus one standard deviation value, color coded by the classication shown in the middle panel (i.e., BPT+WH$\alpha$). Those spaxels with H$\alpha$ flux with a signal-to-noise below 3 or any of the remaining emission lines involved in the BPT diagram below 1 are marked as unknown (dark-red color). Contours are the same already shown in the top-left panel. {\it Bottom-left panel:} distribution along the WHaD diagram, EW(H$\alpha$) vs. $\sigma_{\rm H\alpha}$, for the individual spaxels of the same datacube shown in the previous panels, color coded according to the classification of the ionizing source based on the location within this diagram. {\it Bottom-middle panel:} diagnostic diagram similar to the one already shown in top-middle panel, with the individual spaxels color coded following the new classification based on the WHaD diagram (bottom-left panel). {\it Bottom-right panel:} spatial distribution of the ionized gas similar to that shown in the top-right panel, this time color-coded with the classification obtained using the WHaD diagram shown in the bottom-left. The similarities and differences between the two classiciation schemes are evident when comparing top- and bottom-right panels.}\label{fig:NGC5947}%
\end{figure*}
    %

   \begin{figure*}
   \centering
   \minipage{0.99\textwidth}
   \includegraphics[width=\linewidth,trim={20 20 20 20},clip]{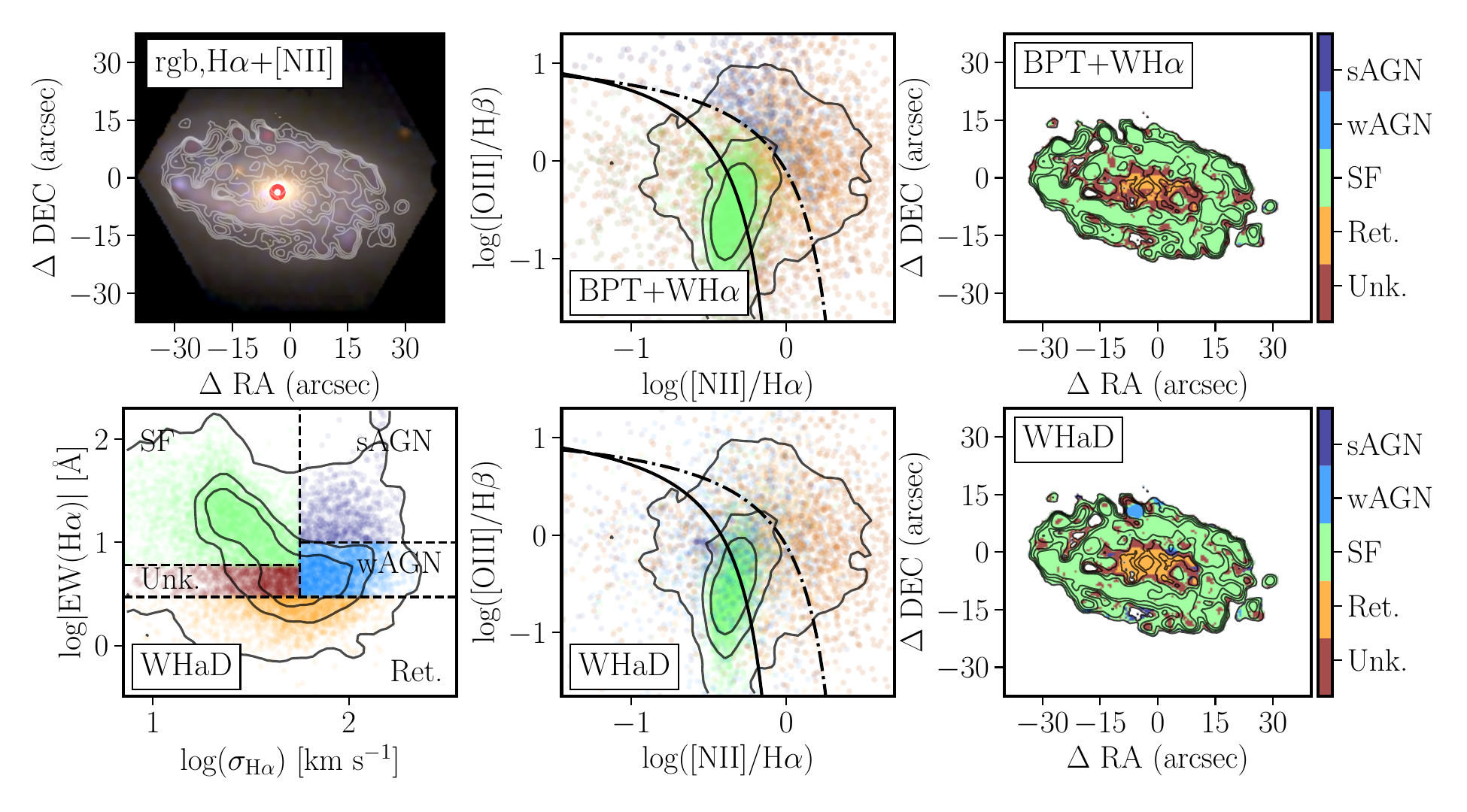}
 \endminipage
 \caption{Similar to Fig. \ref{fig:NGC5947}, corresponding to the eCALIFA data of NGC\,2906} \label{fig:NGC2906}%
    \end{figure*}

    \subsection{BPT-classified ionizing sources across the EW(H$\alpha$)-$\sigma_{\rm H\alpha}$  (WHaD) diagram}
    \label{sec:new_BPT}

\begin{table}
\caption{Results of the 2D Kolmogorov-Smirnov test}
\label{tab:KS}
\begin{tabular}{l|cc}\hline
  \backslashbox{Samples}{Method}  &  BPT+WH$\alpha$ & WHaD \\
  \hline
  SF vs RG           & 0.752 &  0.953 \\
  SF vs O-AGNs       & 0.981 &  0.946 \\
  SF vs X-AGNs       & 0.877 &  0.961 \\
  RG vs O-AGNs       & 0.495 &  0.927 \\
  RG vs X-AGNs       & 0.311 &  0.578 \\
  O-AGNs vs. X-AGNs  & 0.261 &  0.426 \\
\hline
\end{tabular}
\end{table}
    
    We explore now the ability to classify the ionizing source using
    just the EW(H$\alpha$), that has been proved to be crucial to
    distinguish between RG and other ionizing sources, and the
    velocity dispersion \citep[$\sigma_{\rm vel}$,
    following][]{dagos19,carlos20,law21,john22}. Figure
    \ref{fig:diag}, central panel, show the distribution of the
    EW(H$\alpha$) as a function of $\sigma_{\rm H\alpha}$, hereafter
    WHaD diagram, for the same dataset included in the BPT diagram
    (left-side diagram).  It is clear that that three categories of
    ionizing sources selected using the BPT+WH$\alpha$ are located in
    three distiguishable regions.  SF galaxies are all restricted to a
    narrow range of low velocity dispersions and high
    EW(H$\alpha$). Indeed, as already noticed by \citet{law21}, there
    is a very low number of SF galaxies with a velocity dispersion
    above $>$50 \kms ($\sim$2\%), with most of them located in a
    regime $\sim$25 \kms. On the other hand, RG ocupy the regime of
    low EW(H$\alpha$) by construction. In addition, they cover a wider
    range of velocity dispersions, from $\sim$30 \kms to $\sim$300
    \kms, with an average value $\sim$90 \kms (slightly lower for
    eCALIFA than for MaNGA). Finally, the O-AGNs are limited to a
    regime of high EW(H$\alpha$) by construction, covering an even
    wider $\sigma_{\rm H\alpha}$range, with larger values for MaNGA
    than for eCALIFA. Regarding the X-AGNs we do not have the required
    information to replicate the distribution shown in the BPT diagram
    for the \citet{agos19} dataset, however, the sample studied by
    \citet{nata23} distributed across the entire diagram, with a
    preference to larger velocity dispersions than the SF sample.

    We quantify the ability to segregate the different ionizing
    sources using the WHaD diagram in comparison with the BPT one by
    performing a 2D Kolmogorov-Smirnov (KS) test (using the {\sc
      ndtest} python
    package\footnotetext{\url{https://github.com/syrte/ndtest}}).
    Table \ref{tab:KS} lists the values of the KS statistics obtained
    by comparing the distributions of the different ionizing
    subsamples selected based on the BPT+WH$\alpha$ criteria applied
    to the MaNGA dataset (SF,RG and O-AGNs) in addition to the
    \citet{agos19} sample (X-AGNs) across the O3-N2 BPT and the WHaD
    diagrams. We adopted these subsamples as they are the ones with the largest
    number of objects (i.e., a better statistical coverage of the explored
    parameters). The result of this test indicates that the new diagram
    segregates the different subgroups in a more efficient way.
    
    Based on the described distributions and the results of the KS-test, we
    define five regimes in the WHaD diagram.  First, using same
    criteria introduced by \citet{cid-fernandes10} when introducing
    the WHaN diagram, and adopted in the BPT+WH$\alpha$ scheme, we
    classify as RG those sources with an EW(H$\alpha$)$<$3 $\AA$. Then
    we classify as SF those sources with a EW(H$\alpha$)$>$6 $\AA$
    \citep[following][]{sanchez14} and $\sigma_{\rm H\alpha}<$57 \kms,
    a region that comprises 98\% (90\%) of the eCALIFA (MANGA)
    previously selected SF galaxies.  In addition, we define as O-AGNs
    those sources with a EW(H$\alpha$)$>$3 $\AA$ and a
    $\sigma_{\rm H\alpha}>$57 \kms, distinguishing between weak-AGNs
    (wAGN) and strong-AGNs based on the EW(H$\alpha$), using 10$\AA$
    as a boundary between both regimes.  Based on these criteria 86\%
    (95\%) of the O-AGNs selected using the BPT+WH$\alpha$ scheme
    would remain as O-AGNs, and 14\% (1\%) would be reassigned as SF
    in the case of the eCALIFA (MaNGA) sample. Note that the boundary
    in the velocity dispersion was empirically selected to maximize
    (minimize) the agreement (disagreement) between the
    classifications provided by the classical BPT diagram and the new
    proposed WHaD diagram for the optically selected SF and O-AGNs.

    %
    
    If we consider that the BPT-WH$\alpha$ scheme does actually
    reproduce the real distribution of ionizing sources, the new
    classification based on the WHaD diagram would be 100\% reliable
    for the RG (as it is essentially the same scheme), between a
    90-98\% reliable for SF galaxies, and just a 86-95\% reliable for
    O-AGNs. However, we should keep in mind that the initial scheme is
    based on some pre-conceptions, like the fact that AGNs should
    present strong emission lines produced by a hard ionization (i.e.,
    high values of both O3 and N2). As indicated before this is not
    always the case. Figure \ref{fig:diag}, left- and central-panels,
    shows that bona-fide X-AGNs may present a wide range of O3-N2 line
    ratios and an equaly wide range of EW(H$\alpha$) values. Regarding
    X-AGNs, we already established that only 17-34\% of those studied
    by \citet{nata23} and \citet{agos19} are classified as O-AGNs
    using the BPT+WH$\alpha$ scheme.This
    fraction increases to a 50-53\% when using the newly introduced
    WHaD classification procedure. This particularly relevant, as
    X-AGNs are the only sources for which the ionization source is
    fully determined. Based on these results, we should not consider
    that a priori the BPT-WH$\alpha$ offers a better classification
    scheme than the currently proposed one.


    \subsection{Classifying the ionization using the WHaD diagram}
    \label{sec:EW_disp}
    
    Based on this exploration we re-classified the eCALIFA, MaNGA
    galaxies and the X-AGN samples studied in the previous sections
    using the new WHaD diagram and the boundaries proposed before.
    Figure \ref{fig:diag}, right-panel, shows the distribution along
    the O3-N2 BPT diagram based on this new classification, using the
    same nomenclature as the one adopted in the left- and
    central-panels. As a first result, we highlight that the
    distribution for RGs is the same.  Regarding the SF galaxies,
    besides the increase in the number of galaxies (plus 8\%), the
    distributions are very similar.  Only $\sim$1\% of the newly
    classified SF galaxies would be classified as AGNs using the
    BPT+WH$\alpha$ scheme. The main difference is found for the
    O-AGNs, that now are not restricted to the upper-end of the
    right-branch in the BPT diagram. Indeed, they follow a
    distribution more similar to that of the RG or the X-AGNs in this
    diagram. About 27\% (50\%) of the eCALIFA (MaNGA)
    galaxies classified as O-AGNs with the new WHaD scheme would be SF
    based on the BPT+WH$\alpha$ method.  Although this could be due to a
    missclassification introduced by the new procedure, we
    consider more probable that we have disclosed a larger
    population of AGNs thanks to the new method.


    As a final remark, we stress that the results for both
    datasets, eCALIFA and MaNGA, are remarkable similar despite of the
    reported differences between the two datasets, outlined in
    Sec. \ref{sec:data}. The major differences are found in the fraction
    of previously selected O-AGNs (SF) that are not classified as SF (O-AGNs),
    in which both samples disagree by $\sim$20\%. In this regard, we 
    remind that both samples have different selection functions and so far
    we have not applied any correction for the selection function. Thus,
    the reported differences could be just due to the different sample
    selections.


   \begin{figure*}
   \centering
   \minipage{0.99\textwidth}
   \includegraphics[width=\linewidth,trim={20 20 20 20},clip]{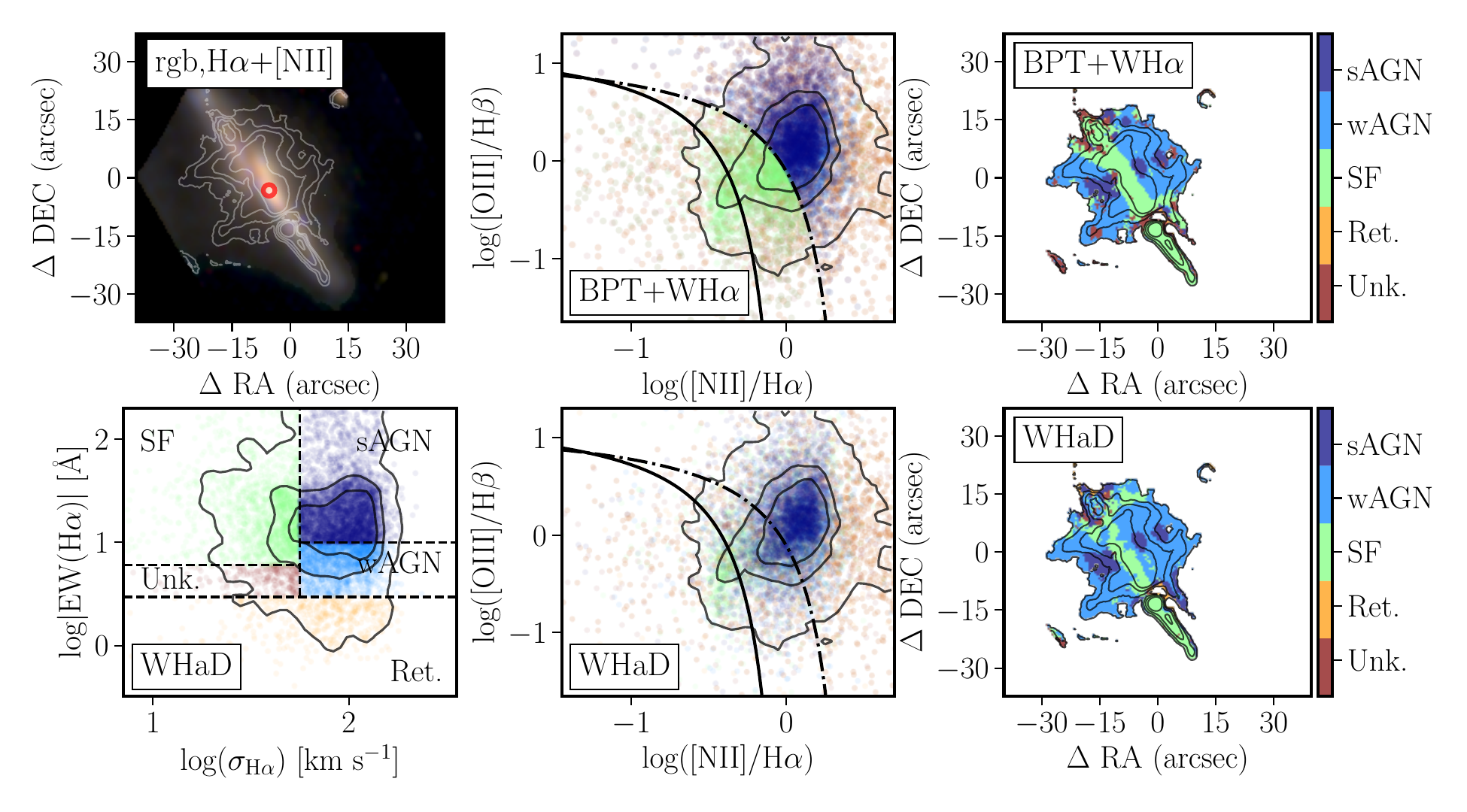}
 \endminipage
   \caption{Similar to Fig. \ref{fig:NGC5947}, corresponding to the eCALIFA data of NGC\,6286.} \label{fig:NGC6286}%
    \end{figure*}

\subsection{Spatial distribution of the ionizing sources using the WHaD diagram}
\label{sec:res_BPT}

%
%

In previous sections we describe how the dominant ionizing source is
determined, for different galaxies (and apertures within them), using a
rather classical/well-stablished approach (Sec. \ref{sec:old_BPT}) in
comparison with the new scheme proposed in this study
(Sec. \ref{sec:new_BPT}-\ref{sec:EW_disp}). As already indicated in
the introduction, and discussed in several previous studies
\citep[e.g.][and references therein]{ARAA,sanchez21}, the ionization
is a process that happens at scales smaller than the ingrated galaxy scale discussed in the previous sections. Like in the case of the
integrated or aperture limited spectra \citep[e.g.][]{kauff03},
diagnostic diagrams and/or a combination of them with additional
parameters is a well established procedure to identify the different
ionizing sources that may be present in a galaxy when using spatial
resolved spectroscopy data \citep[e.g.][for citing just a
few]{law21,john22,eDR2}. To demonstrate that the WHaD diagram provides
with a spatial resolved classification of the ionizing source as good
or even better than the one provided by the BPT+WHa scheme,
we have applied both methods to the spatial resolved dataproducts
derived using \pyp for the full IFS dataset provided by the last
eCALIFA data relase \citep[][]{eDR,eDR2}, already described in
Sec. \ref{sec:data}.  We distribute the outcome of this analysis
through the web
\footnote{\url{http://ifs.astroscu.unam.mx/CALIFA/V500/v2.3/new_diag/}},
describing here in detail the results derived for four cherry-peaked
galaxies that illustrate the general behavior.

Figure \ref{fig:NGC5947} shows the comparison between the spatial
resolved distribution of the ionizing sources identified based on the
BPT+WH$\alpha$ scheme and the WHaD diagram for NGC\,5947. This is a low
inclination early spiral, with a morphology (SBbc) and stellar mass
($\sim$10$^{10.5}$M$_\odot$), very similar to that of the Milky-Way
\citep[indeed, it hosts a solar-neighborhood analog region too;
][]{SNA22}. Figure \ref{fig:NGC5947}, top-left panel, shows an optical
three-color image extracted from the datacube, that traces the light
distribution from the stellar continuum. Contours trace the
distribution the combined flux itensities of H$\alpha$ and \nii. As
expected, the regions with stronger ionized-gas emission trace the
location of the spiral arms, although there is detectable ionized gas
across most of the optical extension of the galaxy. The distribution
across the \oiii\ vs. \nii\ diagnostic diagram for the individual spaxels
is included in the the top-middle panel, color-coded according to the
BPT+WHa classification scheme (Sec. \ref{sec:old_BPT}), with the
spatial distribution of the different ionizing sources shown in the
top-right panel. In this particular implementation of the
classification scheme, we considered that the ionizing source cannot be
correctly identified if the S/N ratio of \Ha\ is lower than three, and
below one for the remaining emission lines involved in this
diagram. Note that for the identification of retired regions, this
criteria could be relaxed, adopting just a cut in EW(H$\alpha$).
As expected, due to the nature of this galaxy, most of the optical
extension is dominated by ionization related to recent SF. Only a
small annular circumnuclear region of $\sim$5$\arcsec$ present an
ionization classified as retired (or unkown). Finally, based
on this scheme the ionization in the central region ($<$3$\arcsec$) is
classified as SF. 

The distribution of spaxels along the WHaD and BPT diagrams for
NGC\,5947 are included in bottom-left and bottom-central panels of
Fig. \ref{fig:NGC5947}, color coded according to the new
classification scheme (Sec. \ref{sec:EW_disp}), with the spatial
distribution of the new classified ionizing sources shown in
bottom-right panel. Despite of the fact that the distribution across
the BPT diagram looks very different, the spatial distribution of the
different ionizing sources is rather similar. Most of the spaxel are
classified as SF, covering a similar region as the one covered by the
previous classifiation, roughly corresponding to the disk of the
galaxy. The annular region around the center is equally classified as
retired (or uknown). However, contrary to the previous classifiation
the ionization in the central region is classified as wAGN/sAGN. Based
on the results discussed in previous sections we cannot conclude that
this is a miss-classification, as up to $\sim$20\% of bona-fide AGNs
(e.g., X-AGNs) could present line ratios compatible with SF, even when
the spectroscopic information is restricted to the central apertures
\citep[e.g.][]{agos19,nata23}. On the other hand, we cannot neither
exclude the possibility that this is a missclassification introduced
by the new method too. . So far, the presence of an AGN has not been
confirmed in NGC\,5947 using non-optical observations.

Figure \ref{fig:NGC2906} presents similar plots for NGC\,2906. This is
a late type, almost face-on, spiral galaxy slightly less massive than
NGC\,5947 (M$_\star\sim$10$^{10}$M$_\odot$). Like in the previous case,
the ionization is classified as SF for most of the regions, in
particular, for those in the galaxy disk. However, contrary to the
previous galaxy, all the ionization in the central regions
($<$10$\arcsec$) is classified as retired (or unknow). The two
classification schemes (BPT+WHa and WHaD) provide very similar
results.  Thus, the spatial distributions of the ionizing sources
included in top- and bottom-left panels of Fig. \ref{fig:NGC2906} are
almost undistiguisable. However, when looking in detail, the WHaD
scheme classifies as wAGN/sAGN a circular region region
$\sim$12$\arcsec$ North and $\sim$5$\arcsec$ East from the center of
the galaxy. On the contrary, based on the BPT+WHa method, the
ionization at this location would be classified as SF. Like in the
previous case we could consider that this is missclassification of the
new method, as obviously an off-center AGN is unlikely.
However, it is known that a supernova exploded in this galaxy at this
exact location before the observing run \citep[SN
2005ip,][]{2005IAUC.8628....1L}, and where a supernovae remnant has been
identified (SNR; Mart\'inez-Rodr\'iguez in prep.). We must recall that
AGN ionization cannot be distinguished from other sources of
ionization such as shocks or SNR based only in the exploration of the
BPT diagram, and additional parameters have to be introduced
\citep[e.g.,][, and in the introduction]{carlos20,cid21}. However, the
WHaD diagram seems to pinpoint SNR that are well below the
classical demarcation lines adopted to separate SF regions from other
sources of ionization.

   \begin{figure*}
   \centering
   \minipage{0.99\textwidth}
   \includegraphics[width=\linewidth,trim={20 20 20 20},clip]{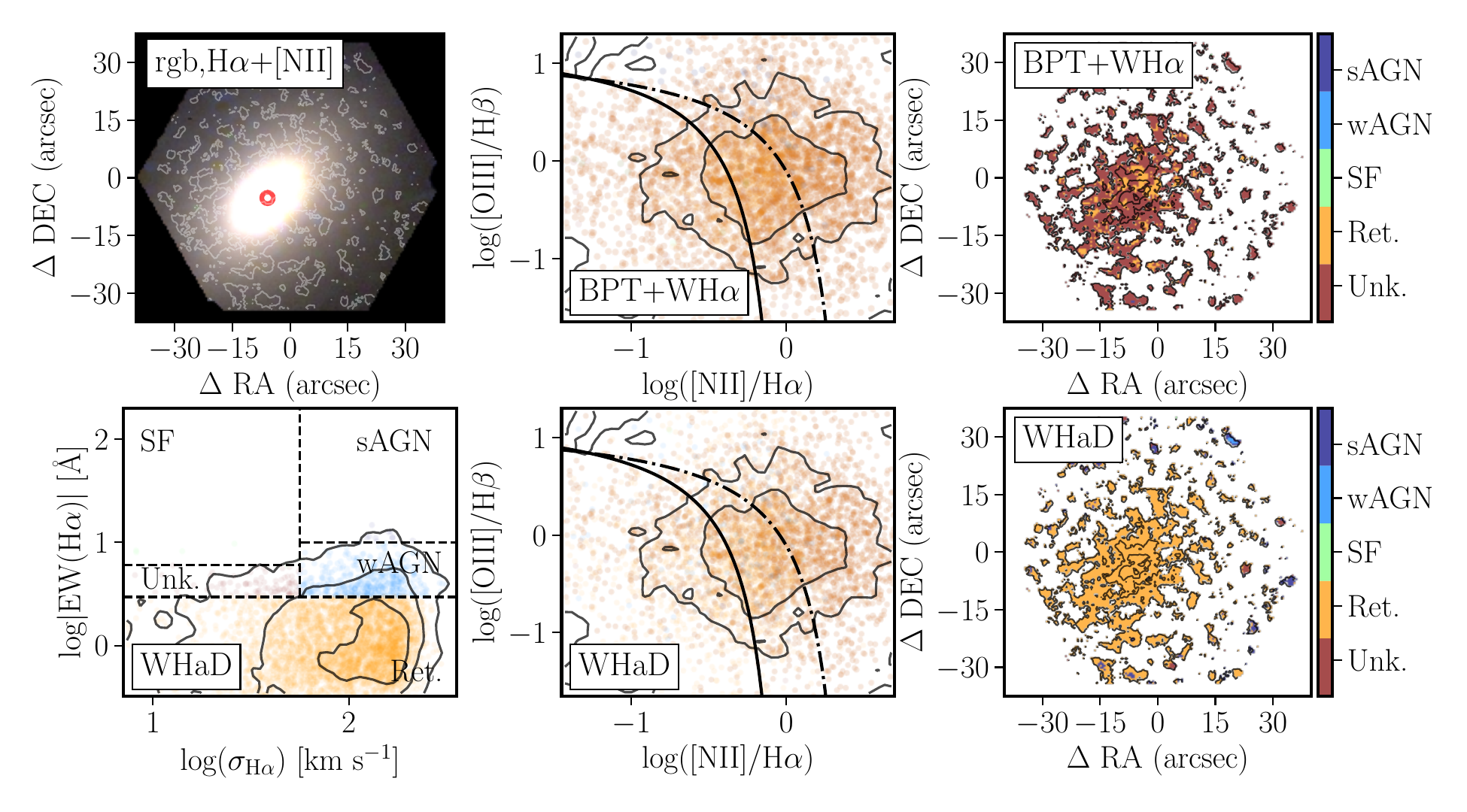}
 \endminipage
   \caption{Similar to Fig. \ref{fig:NGC5947}, corresponding to the eCALIFA data of NGC\,3610} \label{fig:NGC3610}%
    \end{figure*}

Motivated by the previous result we explore how the classifications
based on the two methods comapare in the presence of evident shock
ionization. Figure \ref{fig:NGC6286} shows the same plots already
included in Fig. \ref{fig:NGC5947} and \ref{fig:NGC2906} for the
edge-on spiral galaxy NGC\,6286. This is a luminous infrarred galaxy
\citep{howe10} that presents a well studied galactic outflow
\citep[e.g.][, and references therein]{carlos19}. This outflow is
easily identified in the H$\alpha$+\nii flux intensity contours
included in the top-left panel of Fig. \ref{fig:NGC6286}, having the
archetypical biconical shape. The origin of this galactic outflow is
under discussion. Based on pure optical spectroscopic data, following
the criteria defined by \citet{sharp10}, \citet{carlos19} determined
that this outflows is most probably driven by a strong nuclear SF
process. However, it known that this object host an obscured AGN
\citep{ricci16}, that has been uncovered using X-ray
observations. Irrespectively of the physical origin of the galactic
wind that generated the outflow, the gas along its biconical structure
is clearly ionized by shocks. This is appreciated in top-central and
top-right panels of Fig. \ref{fig:NGC6286}: using the BPT+WHa criteria
the ionized gas at the location of the outflows is classified as
wAGN/sAGN.  Note, once again that this is what it is expected for a
shock-ionized gas.  In addition, the gas along the edge-on disk is
mostly ionized by a SF process. When applying the new classification
scheme (bottom panels), we find a very similar result, with subtle
differences (for instance, the width of the regime classified as SF
along the edge-on disk is slightly narrower when using the new
scheme). We performed a visual inspection of all the galaxies with
detected outflows within the CALIFA sample \citep{carlos19}, finding
similar consistent results. In summary, using the WHaD diagram it is
feasible to identify shock-ionized galacitc outflows as well as using
the BPT+WHa scheme.


Finally, Figure \ref{fig:NGC3610} is similar to
Fig. \ref{fig:NGC5947}-\ref{fig:NGC6286}, for an archeatypical retired
galaxy, NGC\,3610. This is an intermediate mass
(M$_\star\sim$10$^{10.3}$M$_\odot$) elliptical galaxy, with a
weak/diffuse ionized gas emission distributed across most of the FoV
of the IFU data (top-left panel of Fig. \ref{fig:NGC3610}). It
presents a weak and soft X-ray emission that is not compatible with
the presence of an AGN \citep{fabb95}. Despite of the nature of this object (a true
elliptical galaxy or a an early-spiral with a very low
dist-to-bulge ratio), its ionized gas is clearly classified as retired
using the BPT+WHa (top-central and top-right panels of
Fig. \ref{fig:NGC3610}). Note that a substantial fraction of the
spaxels with detected ionized gas would be classified as unknown if a
cut in the S/N ratio of all the lines involved in the BPT diagram is
considered. Using just the EW(H$\alpha$) they would be classified as
retired, being most probably ionized by hot evolved low mass stars in
the post-AGB phases. By construction the WHaD diagram performs a
similar classification, being the difference that most of the
regions which ionizing source is labelled as unknown when using the
full dataset required to explore the BPT diagram are known directly
classified as retired.

In summary we have illustrated how the use of the WHaD classification
scheme, based on one single emission line, provides a reliable
classification of ionizing source for the spatial resolved ionized gas
in galaxies. This classification is compatible, in general, with the
one provided by the BPT+WHa scheme. The reported differences may
reflect possible missclassification cases (e.g., central AGN in
NGC\,5947) or the uncover of previously undetected/missed ionizing
sources (e.g., central AGN in NGC\,5947 again, and the SNR in
NGC\,2906).

\section{Conclusions}
\label{sec:con}

We motivate this work discussing the problems to identify the ionizing
source in galaxies based on the most commonly used method. This method
uses the O3 vs. N2 BPT diagram, identifying regions in which different
ionizing sources are dominant or more frequently observed. The method
present intrinsic problems, due to (i) the S/N requirements and the
wide range of relative fluxes between the involved emission lines, and
(ii) the confussion among regimes in which different ionizing source
could reproduce the observed line ratios.  While the later problem has
been addressed in the literature by the inclussion of additional
parameters, such as the EW(H$\alpha$) and/or $\sigma_{\rm vel}$, to
overcome these problems, the former one cannot be so eassily
addressed.

We have proposed an new method that explores
the location of different ionizing sources in a new diagram (WHaD),
that compares EW(H$\alpha$) and $\sigma_{\rm
  vel}$. This diagram has the virtue of using just one single emission
line, typically the strongest one in the optical range for
all ionizing sources. We use IFU data from eCALIFA and
MaNGA in combination with literature data, to explore the
location of different ionizing sources classified using the standard
procedure (BPT+WHa method) and bona-fide X-ray selected AGNs. Based on
that exploration we define different areas in which the ionizing
source could be classified as: (i) SF, ionization due to young-massive
OB stars, related to recent star-formation activity; (ii) sAGNs/wAGNs,
ionization due to strong (weak) AGNs, and other sources of ionizations
like high velocity shocks; and (iii) RG, ionization due to
hot old low-mass evolved stars (post-AGBs), associated with retired
galaxies and regiones within galaxies (regions in which there is no
star-formation).

We applied the new classification to (i) the same dataset adopted to
define the method and to (ii) the full dataset of spatial resolved
spectroscopic data provided by the eCALIFA survey, comparing the
results provided by the two methods (BPT+WHa vs. WHaD). We found that:

   \begin{enumerate}

   \item Both methods provides exactly the same classification for
     retired regions and galaxies, when only EW(H$\alpha$) is
     considered. If the full set of emission lines required to explore
     the BPT diagram is used the new method recovers a larger
     number of galaxies/regions that cannot be classified using the
     BPT+WHa method.

   \item 90\% (86-95\%) of the galaxies classified as SF (O-AGNs) using the BPT+WHa method would be equally classified using the WHaD diagram.

   \item Around 99\% (50-73\%) of the galaxies classified as SF (O-AGNs) using the WHaD diagram were equally classified using the BPT+WHa method.

   \item Around  50\% of the X-AGNs have been classified as O-AGNs
     using the WHaD method, a significantly larger fraction than the number
     than the one recovered using the more traditional BPT+WHa diagram (17-34\%).

   \item The spatial resolved classification provided by the WHaD is similar to the one provided by the BPT+WHa one for star-forming, retired and high-speed shock ionized regions. However, it increases the number of AGN candidates and AGN-like ionizing sources, that in addition have a more realistic distribution in the BPT diagram that the one shown by O-AGNs.

 \end{enumerate}

 { In summary we consider that the proposed method is an useful tool
   when all the set of emission lines required for other diagnistic diagrams
   are not accesible due to the wavelength range covered by the observations (e.g., in
   Fabry-Perot observations in many cases) or some of them lack sufficient signal-to-noise for
   a proper exploration (e.g., in the case of H$\beta$ for dusty galaxies or regions within galaxies).
   The method has been proved for low-z ($z\sim$0.01) integrated and spatially resolved data. Any
   attempt to apply it at higher redshifts would require a re-evaluation of the method and the proposed
   boundaries, which is beyond the scope of the current exploration.
 }

\begin{acknowledgements}

{ We thanks the anonymous referee for the comments that have improved this manuscript.}
  
S.F.S. thanks the PAPIIT-DGAPA AG100622 project. J.K.B.B. and S.F.S. acknowledge support from the CONACYT grant CF19-39578. JSA acknowledges financial support from the Spanish Ministry of Science and Innovation (MICINN), project PID2019-107408GB-C43
(ESTALLIDOS).

Integral Field Area (CALIFA) survey (http://califa.caha.es/), observed
at the Calar Alto Obsevatory.
Based on
observations collected at the Centro Astron\'omico Hispano Alem\'an
(CAHA) at Calar Alto.
, operated jointly by the Max-Planck-Institut
f\"ur Astronomie and the Instituto de Astrof\'isica de Andaluc\'ia
(CSIC). 

This project makes use of the MaNGA-Pipe3D dataproducts. We thank the
IA-UNAM MaNGA team for creating this catalogue, and the Conacyt
Project CB-285080 for supporting them.

This research made use of
Astropy,\footnote{http://www.astropy.org} a community-developed core
Python package for Astronomy \citep{astropy:2013, astropy:2018}.

Funding for the Sloan Digital Sky 
Survey IV has been provided by the 
Alfred P. Sloan Foundation, the U.S. 
Department of Energy Office of 
Science, and the Participating 
Institutions. 

SDSS-IV acknowledges support and 
resources from the Center for High 
Performance Computing  at the 
University of Utah. The SDSS 
website is www.sdss.org.

SDSS-IV is managed by the 
Astrophysical Research Consortium 
for the Participating Institutions 
of the SDSS Collaboration including 
the Brazilian Participation Group, 
the Carnegie Institution for Science, 
Carnegie Mellon University, Center for 
Astrophysics | Harvard \& 
Smithsonian, the Chilean Participation 
Group, the French Participation Group, 
Instituto de Astrof\'isica de 
Canarias, The Johns Hopkins 
University, Kavli Institute for the 
Physics and Mathematics of the 
Universe (IPMU) / University of 
Tokyo, the Korean Participation Group, 
Lawrence Berkeley National Laboratory, 
Leibniz Institut f\"ur Astrophysik 
Potsdam (AIP),  Max-Planck-Institut 
f\"ur Astronomie (MPIA Heidelberg), 
Max-Planck-Institut f\"ur 
Astrophysik (MPA Garching), 
Max-Planck-Institut f\"ur 
Extraterrestrische Physik (MPE), 
National Astronomical Observatories of 
China, New Mexico State University, 
New York University, University of 
Notre Dame, Observat\'ario 
Nacional / MCTI, The Ohio State 
University, Pennsylvania State 
University, Shanghai 
Astronomical Observatory, United 
Kingdom Participation Group, 
Universidad Nacional Aut\'onoma 
de M\'exico, University of Arizona, 
University of Colorado Boulder, 
University of Oxford, University of 
Portsmouth, University of Utah, 
University of Virginia, University 
of Washington, University of 
Wisconsin, Vanderbilt University, 
and Yale University.

\end{acknowledgements}

%
%
\bibliographystyle{aa}
\bibliography{my_bib} 

%

\end{document}